\documentclass{aastex}
\usepackage{ifthen}

\def \version {_working}

\ifthenelse{\equal{\version}{_apj}}
{
\def \figwidth {0.6 \linewidth}
}
{
\usepackage{emulateapj5}
\usepackage{apjfonts}
\def \figwidth {\linewidth}
}
\usepackage{natbib}

\ifthenelse{\equal{\version}{_working}}
{
}
{
\def \today {January 26, 2003}
}

\usepackage{epsfig}

\makeatletter
\newenvironment{inlinetable}{%
\def\@captype{table}%
\noindent\begin{minipage}{0.999\linewidth}\begin{center}\footnotesize}
{\end{center}\end{minipage}\smallskip}
\newenvironment{inlinefigure}{%
\def\@captype{figure}%
\noindent\begin{minipage}{0.999\linewidth}\begin{center}}
{\end{center}\end{minipage}\smallskip}
\makeatother

\shorttitle{Density profile of dark matter halos}
\shortauthors{Dahle, Hannestad, \& Sommer-Larsen}
\slugcomment{ApJL, in press}

\begin{document}

\title{The density profile of cluster-scale dark matter halos}

\author{H{\aa}kon Dahle, Steen Hannestad}
\affil{NORDITA, Blegdamsvej 17, DK-2100 Copenhagen {\O}, Denmark\\
{\tt dahle@nordita.dk }} 
\author{Jesper Sommer-Larsen}
\affil{Theoretical Astrophysics Center, Juliane Maries Vej 30, DK-2100 Copenhagen {\O}, Denmark}

\begin{abstract} {We measure the average gravitational shear profile of 6 massive clusters ($M_{\rm vir} \sim 10^{15} M_{\sun}$) at $z=0.3$ out to a radius $\sim 2h^{-1}$~Mpc. The measurements are fitted to a generalized NFW-like 
halo model $\rho (r)$ with an arbitrary $r \rightarrow 0$ slope $\alpha$. The data are well fitted by such a model with a central cusp with $\alpha \sim 0.9 - 1.6$ (68\% confidence interval). For the standard-NFW case $\alpha = 1.0$, we find a concentration parameter $c_{\rm vir}$ that is consistent with recent predictions from high-resolution CDM N-body simulations. Our data are also 
well fitted by an isothermal sphere model with a softened core. For this model, our $1\sigma$ upper limit for the core radius corresponds to a limit $\sigma_{\star} \leq 0.1 {\rm cm}^2 {\rm g}^{-1}$ on the elastic collision cross-section in a self-interacting dark matter model. 
}\end{abstract}


\section{Introduction}

Although the cold dark matter (CDM) model of structure formation has been 
successful in explaining many observable properties of the universe, there 
remain some notable discrepancies with observations. 
In particular, numerical simulations of dark matter haloes in 
CDM universes appear to overproduce the abundance of dark matter sub-haloes 
corresponding to dwarf galaxies by at least an order of magnitude (Klypin et al.\ 1999; 
Moore et al.\ 1999). Also, simulations indicate a transfer of angular momentum from the 
baryonic matter component to the non-baryonic dark matter during the assembly of disk galaxies, 
which leads to a conflict with the observed angular momentum properties of disk 
galaxies (e.g., Navarro \& Steinmetz 1997,2000; Sommer-Larsen, G{\" o}tz, \& Portinari 2002).  

Finally, the simulations predict that dark matter halos in a wide mass range from dwarf 
galaxies to massive clusters follow an universal density profile with a central cusp  
$\rho(r) \propto r^{-\alpha}$ with $\alpha$ in the range $1.0 < \alpha < 1.5$
(Navarro, Frenk, \& White 1996, 1997; hereafter collectively NFW; Moore et al.\ 1999). 
This appears to be contradicted by some observations that indicate an almost constant-density core 
rather than a cusp, both on galaxy scales (e.g., Dalcanton \& Bernstein 2000; 
Salucci \& Burkert 2000; de Blok et al.\ 2001) and on cluster scales (Tyson, Kochanski, 
\& Dell'Antonio 1998), although the interpretation 
of these results is controversial (e.g., Broadhurst et al.\ 2000; Shapiro \& Iliev 2000; Czoske et al.\ 2002). This has led to suggestions that the dark matter 
properties may deviate from standard CDM, e.g., the dark matter could be warm 
(Col{\'i}n, Avila-Reese, \& Valenzuela 2000; Sommer-Larsen \& Dolgov 2001), 
repulsive (Goodman 2000), fluid (Peebles 2000), fuzzy (Hu, Barkana, \& Gruzinov 2000), 
decaying (Cen 2001), annihilating (Kaplinghat, Knox, \& Turner 2000), self-interacting 
(Spergel \& Steinhardt 2000; Yoshida et al.\ 2000, Dav{\'e} et al.\ 2001) or both warm and 
self-interacting (Hannestad \& Scherrer 2000). 
Alternatively, it has been suggested that stellar feedback from the first generation of stars formed 
in galaxies was so efficient that the remaining gas was expelled on a timescale comparable to, or less than, the local dynamical timescale. The dark matter subsequently adjusted to form an approximately 
constant density core (e.g., Gelato \& Sommer-Larsen 1999). This is however unlikely to affect cluster
cusps.   

It is dangerous to draw 
wide-ranging conclusions about halo density profiles from a single object, 
even when based on high-quality data from strong gravitational lensing. Clearly, a larger number 
of clusters should be investigated using a range of methods.  
Other studies have sought to put constraints on $\alpha$ from the observed abundances of 
strongly lensed arcs produced by clusters (Meneghetti et al.\ 2001; 
Molikawa \& Hattori 2001; Oguri, Taruya, \& Suto 2001), and find best fit values of $\alpha$ 
similar to CDM predictions.

Previous weak lensing measurements of cluster mass profiles have usually only considered 
two models, a singular isothermal sphere (SIS) model with $\rho \propto r^{-2}$ and the 
NFW model. Typically, it is found that the NFW model provides a marginally better fit 
to the data, but usually not at a very high significance (Clowe et al.\ 2000; 
Clowe \& Schneider 2001; Hoekstra et al.\ 2002). Many of the clusters studied in this way 
show evidence for substructure, indicating that they may not be in dynamical equilibrium.   

Here, we consider a density profile which is a generalized version of the NFW model, 
as described in \S 2. The data we present here,  
detailed in \S 3, consists of measurements of gravitational shear over a wide range of radii 
around 6 clusters at $z \simeq 0.30$.  
In \S 4 we present the results of our model fits, and in \S 5 we discuss the implications for 
cluster-scale dark matter halos and dark matter physics.   
We consider both a spatially-flat $\Lambda$CDM model with $(\Omega_{0},\lambda_{0}) = (0.3,0.7)$
and a SCDM (Einstein-de Sitter) cosmology with $(\Omega_{0},\lambda_{0}) = (1,0)$, and we use 
$H_0 = 100 h\, {\rm km}\, {\rm s}^{-1} {\rm Mpc}^{-1}$.

\section{Halo model}
\label{sec:model}

Here we consider a generalized halo density profile on the form (Zhao 1996; Jing \& Suto 2000)  

\begin{equation}
\rho(r) = \frac{\delta_c \rho_c}{(r/r_s)^\alpha (1+(r/r_s))^{3-\alpha}} ,   
\label{eq:densprof} 
\end{equation}

\noindent
which corresponds to the standard-NFW model for $\alpha = 1$, but 
has an arbitrary inner slope. Moore et al.\ (1999) favor an inner slope $\alpha = 1.5$ and adopt
a model with a slightly different functional form than we consider here, 
but with the same asymptotic behavior as an $\alpha = 1.5$ model in
equation~(\ref{eq:densprof}) for large and small scales.    

NFW originally defined a concentration parameter $c$ by $c = r_{200}/r_s$, 
where $\overline{\rho}(r_{200}) = 200 \rho_{\rm c}$ and $\rho_{\rm c}$ is the critical density. 
For the model above, the characteristic density $\delta_c$ is 

\begin{equation}
\delta_c = \frac{200}{3} \left[\int_0^1 x^2 (c x)^{-\alpha}(1+c x)^{\alpha-3} dx \right]^{-1}.
\label{eq:deltc} 
\end{equation}

\noindent

Keeton \& Madau (2001) define a slightly different concentration parameter $c_{-2} = r_{\rm vir}/r_{-2}$, 
where $r_{\rm vir}$ is the virial radius of the halo 
and $r_{-2} = (2 - \alpha) r_s$ is the radius at which $d\ln \rho / d\ln r = -2$. 
At the redshift of the clusters we examine $(z=0.3)$, N-body simulations of clusters in a $\Lambda$CDM universe 
indicate that the virial radius is related to $r_{200}$ by $r_{\rm vir} = 1.194 r_{200} + 0.003{\rm Mpc}$ 
(M. G{\"o}tz, 2002, private communication). For the SCDM model we make the approximation 
$r_{\rm vir} \simeq r_{200}$, since $\overline{\rho}(r_{\rm vir}) = 178 \rho_c \simeq 200 \rho_c$.

Bullock et al.\ (2001) use high-resolution N-body simulations to predict the mass- and 
redshift-dependence of the concentration parameter $c_{\rm vir} = r_{\rm vir}/ r_s$ for NFW-type halos. Simulated dark matter halos show a significant dispersion about the median value of the concentration parameter at a given mass (Jing 2000), and Bullock et al.\ (2001) predict a scatter of $\Delta ({\rm log} c_{\rm vir}) = 0.18$. Bullock et al.\ (2001) found the 
redshift-dependence of this median value to be $c_{\rm vir} \propto (1+z)^{-1}$.
For very massive cluster halos ($M_{\rm vir} \sim 10^{15} M_{\sun}$), 
the concentration is low, and $r_s$ is of order several hundred kpc, meaning that the difference between 
various values of $\alpha$ is evident even at cluster radii associated with the weak lensing regime.

\section{Lensing data} 
\label{sec:data}

The cluster sample used here was selected from a larger sample of $38$
clusters targeted for weak lensing mass measurements. The data set, and the 
methods for data reduction and shear estimation have been described 
elsewhere (Dahle et al.\ 2002). In this letter we focus on the  
density profiles of a subset of six clusters of galaxies at $z=0.3$.  
These are the most distant clusters in the sample of Dahle et al.\ (2002)  
for which wide-field imaging data exist, allowing us to probe the cluster 
mass profile beyond the virial radius.  

The data were obtained with the UH8K mosaic CCD camera at the 2.24 m 
University of Hawaii telescope at Mauna Kea Observatory (see Dahle et al.\ 2002 for 
details). The clusters were selected from the X-ray luminous 
cluster samples of Briel \& Henry (1993) and Ebeling et al.\ (1996), and all have X-ray luminosities 
L$_{\rm x, 0.5-2.5 keV} \geq 4 \times 10^{44}$ erg s$^{-1}$. The virial masses $M_{\rm vir}$
of half of the clusters have been estimated by Irgens et al.\ (2002), who find values
 in the range $1.1\times 10^{15} < h M_{\rm vir}/M_{\sun} < 3.3\times 10^{15} $. 

The gravitational shear $\gamma$ and convergence $\kappa$ produced by 
the halo model given in equation~(\ref{eq:densprof}) cannot be expressed in terms of 
simple functions, but we have derived approximate relations that are accurate to about 
2\% for the part of parameter space that we are able to probe with our data.
The observable tangential distortion $g_T$ 
is related to the tangential shear and convergence by $g_T = \gamma_T /(1 - \kappa)$ at all radii larger than the Einstein radius $r_E$. 

As first suggested by Kaiser \& Squires (1993),  
the signal-to-noise ratio of weak lensing observations can be boosted by ``stacking'' the 
weak shear measurements of a number of similar objects. Here, we stack our tangential distortion 
measurements for the clusters 
\objectname{A959}, \objectname{A1351}, \objectname{A1705}, \objectname{A1722}, \objectname{A1995} and \objectname{A2537}, which all lie in the redshift interval $0.285 < z < 0.325$. For each cluster, the average tangential distortion $\langle g_T \rangle$ vs. radius was measured in 24 radial bins, assuming the peak in the reconstructed projected mass density (see Dahle et al.\ 2002) to be the cluster center. 

\ifthenelse{\equal{\version}{_apj}}
{}{
\begin{inlinefigure}
\centering\epsfig{file=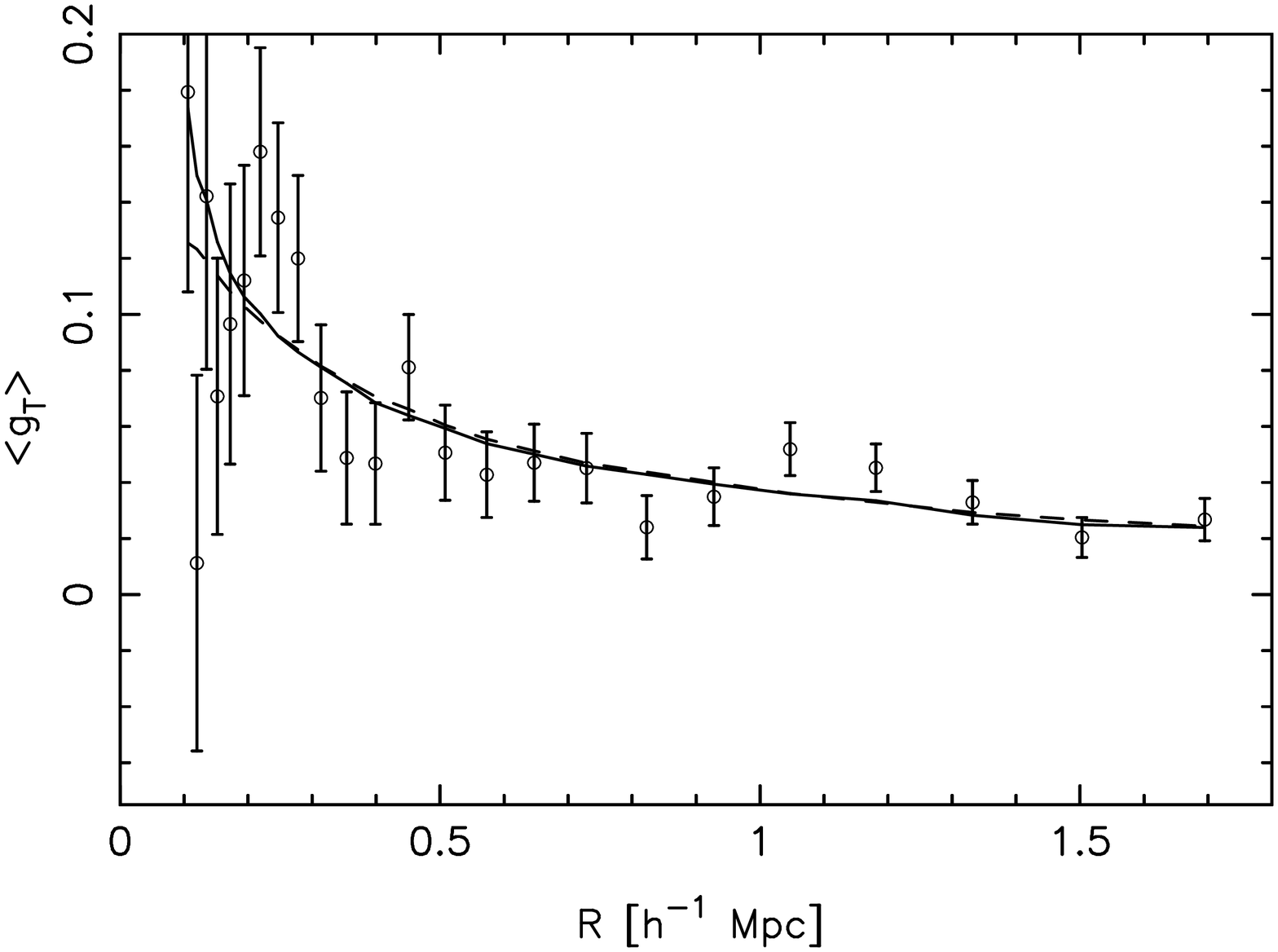,width=\figwidth}
\caption[Fits to radial shear profile.]
{The points with error bars show the averaged gravitational lensing distortion of background galaxies as a function of radius. 
The solid line indicates the best fit generalized-NFW model and the dashed line indicates the best fit NSIS model. Corrections for cluster member contamination have been applied to the model curves. The physical radii plotted here 
are calculated for the SCDM cosmology and would be 15\% larger for the $\Lambda$CDM cosmology.
} 
\label{fig:shearprofilefit}
\end{inlinefigure}
}

We re-scale our clusters to the same mass (represented by a halo with velocity dispersion 
$\sigma_v = 1150 {\rm km~s}^{-1}$, which is the average value for our clusters) using a SIS model fit to the observed tangential distortion. In doing this, we neglect the mass-dependence 
of the concentration parameter and the $c_{\rm vir} \propto (z+1)^{-1}$ redshift dependence 
predicted by Bullock et al.\ (2001). These are both only minor corrections, 
since the clusters we consider here have a very small spread in redshift and modest scatter in mass, 
and at this point we do not want to impose additional constraints on our model parameters by assuming scaling laws like those predicted by NFW or by Bullock et al.\ (2001).  We also scale the clusters to a uniform lens redshift $z = 0.3$ by taking into account the variation in lensing strength and angular diameter with redshift. 

The faint galaxy catalogs which were used to estimate $\langle g_T \rangle$ were generated by selecting  
galaxies that were detected at low significance $\nu$ ($6 < \nu < 100$).  
These galaxies have typical redshifts $z_{bg} \sim 0.9$ and typical magnitudes $V=24$, $I=23$. At small projected radii from the cluster center there will be a significant fraction of cluster galaxies in the faint galaxy catalogs, and the shear estimates will be systematically biased towards lower values due to this contamination. The 
importance of this effect can be estimated by assuming 
that there is negligible contribution by cluster galaxies at the edges of the UH8K fields (i.e., beyond 
$\sim 1.5 h^{-1}$~Mpc). Thus, the true density of background galaxies can be estimated, and the cluster 
contribution to the faint galaxy density can be calculated in azimuthally averaged bins as a function of 
radius from the cluster center. By calculating the average contamination for the 
six clusters that go into the stack, we can significantly reduce the effects 
of cluster richness variations, substructure and background galaxy density fluctuations. 

\section{Results} 
\label{sec:results}

The average tangential distortion profile in a SCDM universe is shown in Figure~\ref{fig:shearprofilefit} along with 
the best fit generalized-NFW model curve. The $68\%$ confidence interval for each parameter is given in 
Table~\ref{tab:parvalues} for the two cosmologies we consider. The observed profile is well represented by equation~(\ref{eq:densprof}) for both the SCDM and $\Lambda$CDM case, with best fit $\chi^{2}/d.o.f. = 23.66/21$ and $22.78/21$, respectively. 

\ifthenelse{\equal{\version}{_apj}}
{}{
\begin{inlinefigure}
\centering\epsfig{file=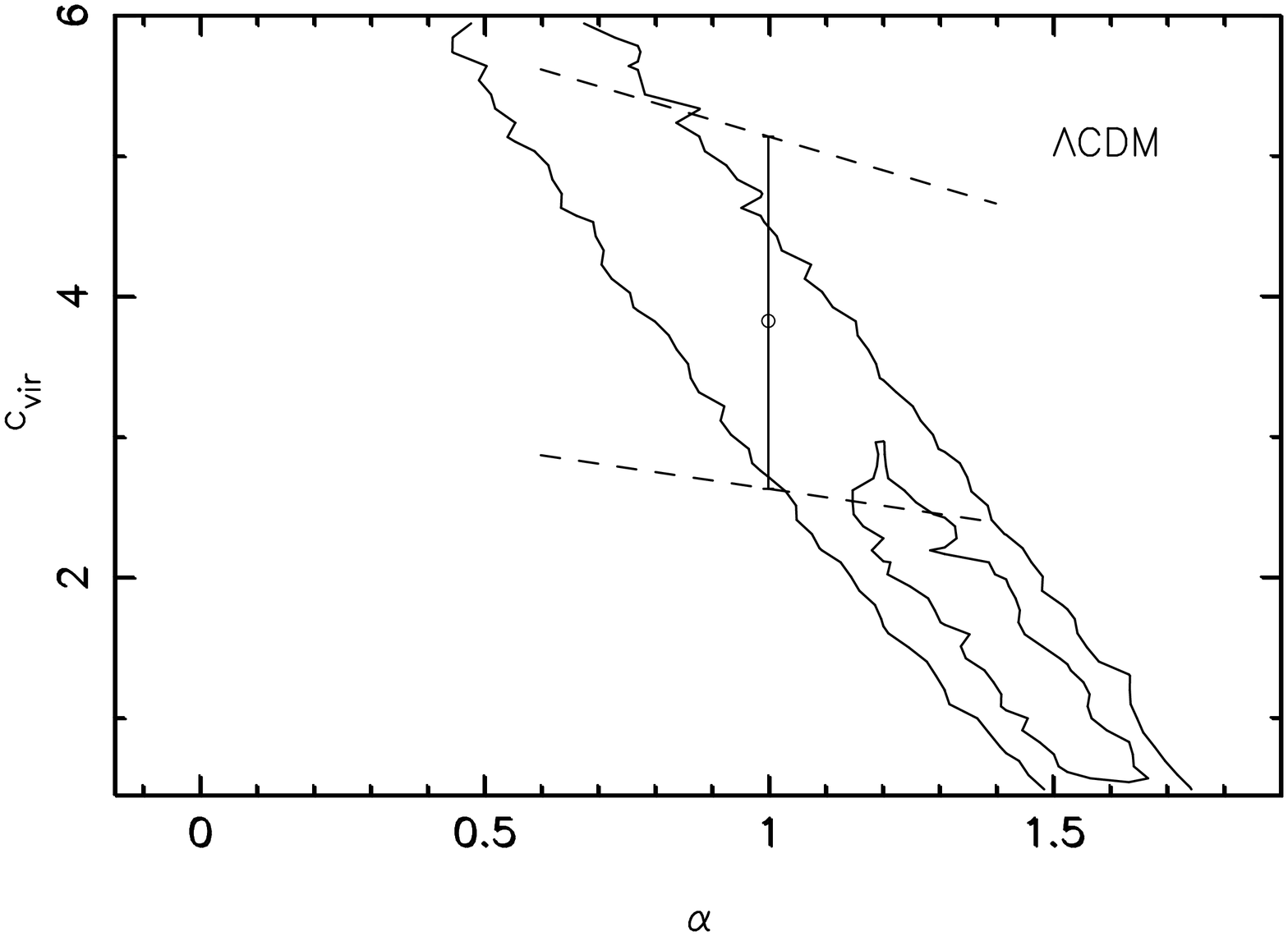,width=\figwidth}
\caption[C-alpha joint confidence intervals.]
{The contours indicate the $68\%$ and $95\%$ joint confidence intervals on $c_{\rm vir}$ and $\alpha$, derived from variations in $\chi^2$ with respect to the minimum value.  Also shown is the mean value and scatter in the concentration parameter for a $M_{\rm vir} = 2 \times 10^{15} M_{\sun}$ NFW halo predicted by Bullock et al.\ (2001) for a $\Lambda$CDM cosmology. The dashed lines indicate lines along which the two parameters are degenerate.} 
\label{fig:confint_C_alpha_lambda}
\end{inlinefigure}
}

A massive central galaxy can significantly contribute to the projected 
mass density at small radii. Williams, Navarro \& Bartelmann (1999) found that 
strong lensing constraints on clusters with $\sigma_v \sim 1500-2000 {\rm km~s}^{-1}$
are consistent with a standard-NFW profile, but $\sigma_v \sim 1000 {\rm km~s}^{-1}$ clusters are inconsistent 
with this model, unless the central surface density is significantly enhanced by the central galaxy. 
We check the robustness of 
our results to the inclusion of such an additional mass component by adopting a 
combined model which is a superposition of the dark matter profile in equation~(\ref{eq:densprof})  
and a SIS mass profile with $\sigma_v = 300 {\rm km~s}^{-1}$. This velocity dispersion is  
a typical value measured for brightest 
cluster galaxies (e.g., Oegerle \& Hoessel 1991) and is consistent with 
the extra central mass of $\sim 3 \times 10^{12} h^{-1} M_{\sun}$ required to 
produce the arcs studied by Williams et al.\ (1999).  

\ifthenelse{\equal{\version}{_apj}}
{}{
\begin{inlinefigure}
\centering\epsfig{file=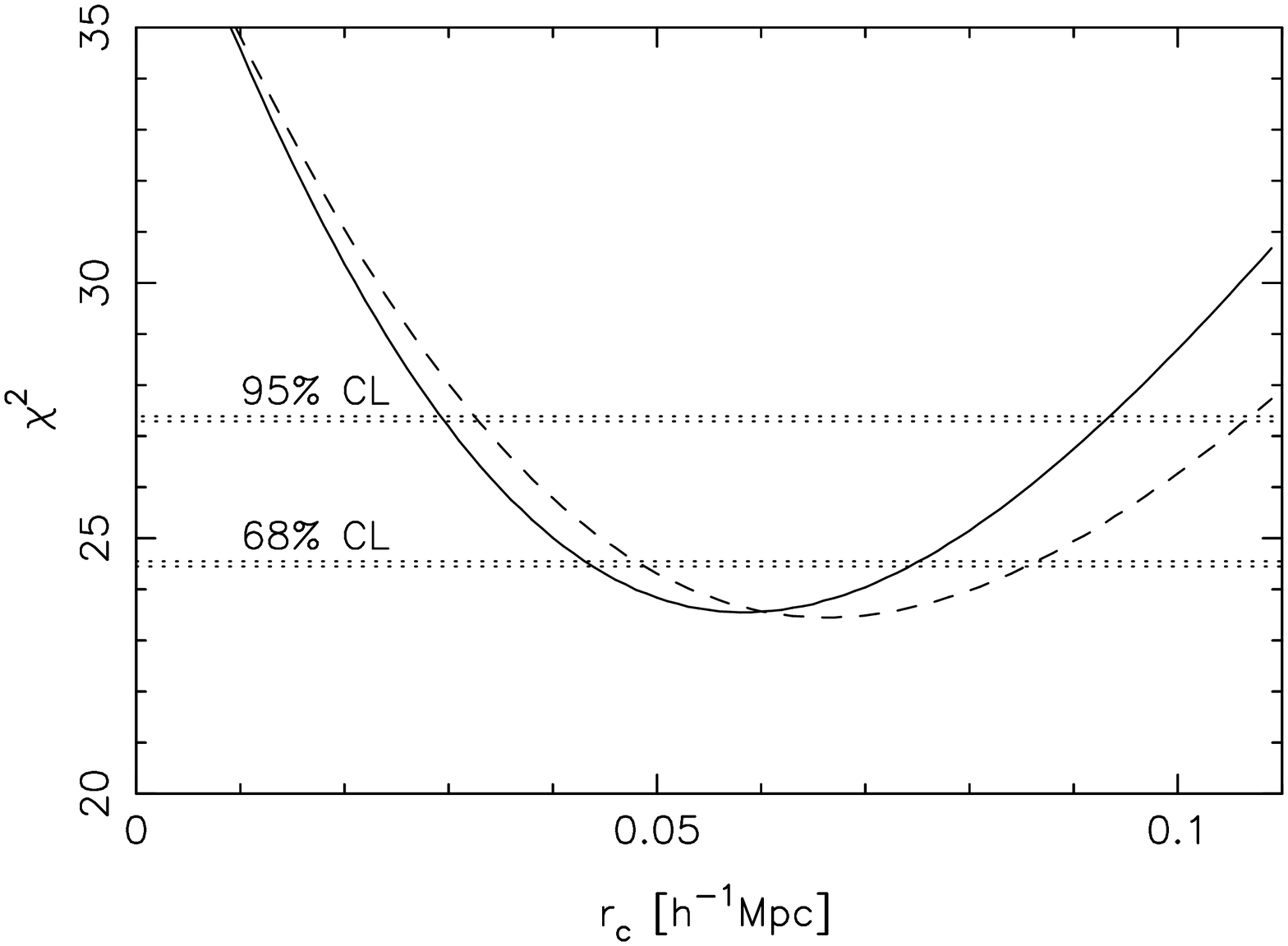,width=\figwidth}
\caption[NSIS fit.]
{Constraints on the core radius $r_c$ for a NSIS profile (solid line: SCDM; dashed line: $\Lambda$CDM).} 
\label{fig:NSIS}
\end{inlinefigure}
}

As shown in Table~\ref{tab:parvalues}, the inclusion of an additional SIS mass 
profile did not significantly alter the best fit parameter values, and 
the quality of the fits was also not strongly affected. Not surprisingly, it is seen that with the 
inclusion of an extra mass component with a steep $r^{-2}$ slope in the center, the 
best-fit inner slope of our generalized NFW model needs to become slightly shallower to 
match the observed shear. Figure~\ref{fig:confint_C_alpha_lambda} shows 2-parameter joint confidence regions 
for $c_{\rm vir}$ and $\alpha$ for the $\Lambda$CDM cosmology. For the $\alpha = 1$ case, the confidence 
intervals for $c_{\rm vir}$ are found to be consistent with the range of 
concentration parameters for $M_{\rm vir} = 2 \times 10^{15} h^{-1} M_{\sun}$ halos 
predicted from N-body simulations by Bullock et al.\ (2001).

\begin{inlinetable}
\caption{Best fit model parameters.}
\begin{tabular}{ccccc}
\multicolumn{5}{c}{SCDM ($\Omega_m = 1$, $\Omega_{\Lambda} = 0$)} \\
\multicolumn{1}{c}{ } &
\multicolumn{2}{c}{Generalized-NFW} &
\multicolumn{2}{c}{Generalized-NFW + SIS} \\
Parameter & Best fit value & $68\%$ conf. int. & Best fit value & $68\%$ conf. int. \\
$c_{\rm vir}$ & 1.3 & $0.2 - 4.2$ & 1.3 & $0.3 - 2.8$ \\
$\alpha$ & 1.5 &  $0.9 - 1.6$ & 1.4 &  $1.1 - 1.6$ \\
$r_{\rm vir}$ & 1.34 &  $1.28 - 1.49$ & 1.34 &  $1.28 - 1.43$ \\
\multicolumn{5}{c}{$\Lambda$CDM ($\Omega_m = 0.3$, $\Omega_{\Lambda} = 0.7$)} \\
\multicolumn{1}{c}{ } &
\multicolumn{2}{c}{Generalized-NFW} &
\multicolumn{2}{c}{Generalized-NFW + SIS} \\
Parameter & Best fit value & $68\%$ conf. int. & Best fit value & $68\%$ conf. int. \\
$c_{\rm vir}$ & 1.7 & $0.4 - 2.6$ & 2.1 & $0.2 - 4.8$ \\
$\alpha$ & 1.4 & $1.3 - 1.6$ & 1.3 & $0.9 - 1.6$ \\
$r_{\rm vir}$ & 1.97 & $1.91 - 2.03$ & 1.91 & $1.79 - 2.10$ \\
\end{tabular}
\label{tab:parvalues}
\end{inlinetable}

We also fitted our data to a non-singular isothermal sphere (NSIS) model on the form 
$\rho(r) = \sigma_v^2/[2 \pi G (r^2 + r_c^2)]$,   
where $r_c$ is the core radius. Our data were also well described 
by this model, with a best fit $\chi^{2}/d.o.f. = 23.5/22$ for both SCDM and 
$\Lambda$CDM. Our best fit values were $r_c = 58h^{-1}$~kpc and $r_c = 66h^{-1}$~kpc for 
SCDM and $\Lambda$CDM, respectively. The best fit model curve is plotted in 
Figure~\ref{fig:shearprofilefit}, and the resulting 68\% and 95\% confidence 
intervals for $r_c$ are shown in Figure~\ref{fig:NSIS}. 

\section{Discussion}

We find that the average density profile of 6 massive clusters is well fitted by a generalized-NFW model 
with central slope $\alpha \sim 0.9-1.6$. This (steep) slope fits
well with predictions from collisionless CDM simulations. The best fit slope of $\alpha = 1.3-1.5$ 
is slightly steeper than the canonical NFW value, although $\alpha=1$ is only disfavored at less than
90\% confidence for any of the models. 
For $\Lambda$CDM, $\alpha \lesssim 0.5$ can be excluded with 95\% confidence.
The inclusion of an additional SIS mass profile corresponding to a massive 
central galaxy only resulted in minor changes for the best fit parameters. 
This indicates that any additional mass component associated with the central galaxy 
does not make a significant contribution to the total projected density on the scales 
that we probe in our weak lensing study. 

Our constraints on $\alpha$ are compatible with recent strong lensing studies 
of massive clusters: Smith et al.\ (2001) estimate  $\alpha = 1.3$ for \objectname{A383} and 
Gavazzi et al.\ (2002) find $0.7 \leq \alpha \leq 1.2$ for \objectname{MS2137-23} (see, however,
Sand, Treu \& Ellis 2002, who find a best fit $\alpha = 0.35$ and 
$\alpha < 0.9$ at 99\% CL for the same cluster).  
Our results are also consistent with recent X-ray cluster studies based on Chandra data: 
Arabadjis, Bautz, \& Garmire (2002) find that the mass profile of \objectname{CL 1358+6245}
is well fit by a standard-NFW $\alpha = 1$ profile, and  Lewis, Buote, \& Stocke (2003) find 
$\alpha = 1.19 \pm 0.04$ for the central slope of \objectname{A2029}. 

We find that an isothermal sphere model with a softened core can also produce a good fit to our data. 
This indicates that our weak lensing data is unable to discriminate between an outer slope 
of $r^{-2}$ and $r^{-3}$. A pure SIS model is however strongly excluded 
at the 99.9\% CL (see Fig.~\ref{fig:NSIS}). 
Yoshida et al.\ (2000) use simulations 
to study the effect of self-interacting dark matter on the structure of cluster halos. 
They find that even a small elastic collision cross-section
of $\sigma_{\star} = 0.1 {\rm cm}^2 {\rm g}^{-1}$   
will significantly affect the central density profile. 
Meneghetti et al.\ (2001) study the 
strong lensing properties of the cluster models of Yoshida et al.\ (2000), 
and find that even a cross-section as small as 
$0.1 {\rm cm}^2 {\rm g}^{-1}$ is incompatible with the 
observed abundances of strongly lensed radial and tangential arcs in clusters. 
Their study is based on a simulated cluster at $z = 0.278$ of 
final virial mass $7.4 \times 10^{14} h^{-1} M_{\sun}$ in a $\Lambda$CDM 
universe. This cluster is at a similar redshift as the clusters in our data 
set, but our clusters are on average about twice as massive. 
From the $r_c$ values tabulated by Meneghetti et al.\ (2001), 
we find that the predicted core radius for
$\sigma_{\star} = 0.1 {\rm cm}^2 {\rm g}^{-1}$ of $r_c = 80h^{-1}$~kpc
 is similar to our $1\sigma$ (and $2\sigma$) upper limit for the 
core radius shown in Figure~\ref{fig:NSIS}. Here, we have taken into account 
a factor $\sqrt{3}$ difference in the definition of $r_c$ and a $M^{1/3}$
scaling of lengths. A cross-section of $1 {\rm cm}^2 {\rm g}^{-1}$ would 
produce a core radius of $180h^{-1}$~kpc, which is strongly 
excluded by our data.  
In the small mean free path limit of ``fluid'' dark matter, a core-collapse 
producing a SIS-type profile is expected. This is  
strongly excluded by our data, and the fluid limit is also excluded 
by the observed ellipticities of clusters (Miralda-Escud{\' e} 2002).   

Dav{\'e} et al.\ (2001) show that cross-sections of order  
$5 {\rm cm}^2 {\rm g}^{-1}$ are needed to produce a good fit to the large 
apparent cores in dwarf galaxies. By introducing a velocity-dependent 
interaction cross-section for the dark matter particles, it may still be 
possible to reconcile our results with a self-interacting dark matter model 
that can produce constant-density cores in dwarf galaxies. 
Even if such a model may still be allowed, its velocity-dependence is already 
strongly confined by other astrophysical constraints, including the 
demographics of supermassive black holes in galaxy nuclei 
(Hennawi \& Ostriker 2001).     
   
It should be noted that even though dark matter models which produce
large constant-density cores on clusters scales appear to be ruled out, 
many of the non-standard dark matter models are easily compatible with 
cusped cluster halos. One example is warm dark matter where the thermal velocity
of the dark matter particles can erase the cusps of dwarf galaxy halos,
but almost not affect more massive halos at all.

\ifthenelse{\equal{\version}{_working}}
{
\bibliographystyle{Apj}
\bibliography{astro,clusters,weaklensing,kaiser_ref,kaiser_nonref}
}
{

}

\end{document}